\documentclass[shortnote,twocolumn]{jpsj2} %% for short notes
\title{Mixing-Demixing transition in 1D boson-fermion mixture at low fermion densities}

\author{\textsc{Yousuke Takeuchi}$^{1}$ and \textsc{Hiroyuki Mori}$^{2}$}

\inst{$^{1}$Department of Quantum Matter, ADSM, Hiroshima University,
 1-3-1 Kagamiyama, Higashi-Hiroshima 739-8530, JAPAN\\
$^{2}$Department of Physics, Tokyo Metropolitan University,
Minamiohsawa 1-1, Hachioji-shi, Tokyo 192-0397, JAPAN\\}

\kword{ultra-cold atomic gases, boson-fermion mixture, mixing-demixing transition, Monte Carlo simulation}

\begin{document}
\maketitle

%\section{Introduction} %% No sections necessary for express letters, letters and short notes
Boson-fermion mixtures realized in recent years with ultra-cold atomic gases have been known as a new many-body system exhibiting exciting phenomena and have attracted attention from both experimentalists and theorists\cite{Schreck}-\cite{Modugno}. The mixture systems are constructed not only in three-dimensional geometry but also in one and two dimensions. Several interesting features have been pointed out theoretically on the one-dimensional (1D) boson-fermion mixtures: Demixing instability\cite{Das, Cazalilla}, density wave\cite{Lewenstein, Mathey}, Mott transition on a lattice\cite{Albus, Roth}, and exact solutions for ground state properties\cite{Imambekov, Batchelor}.

In this short note we shed light on the mixing-demixing transition of 1D boson-fermion mixtures. 
We could expect the fermions and the bosons are demixed with strong-enough repulsive interactions between them. 
According to a mean-field calculation\cite{Das}, the transition to the demixing phase occurs in a system with a low fermion density. On the other hand, a Luttinger liquid analysis revealed the possibility of the mixing-demixing transition when the number densities of the fermions and bosons were largely different\cite{Cazalilla}.
It was also shown in Bethe ansatz solution that the mixture was always in the mixing phase as far as the fermion-boson interaction had the same magnitude as the boson-boson interaction\cite{Imambekov}.  

In our previous work we numerically investigated the mixing-demixing transition of the boson-fermion mixture on a 1D lattice at an incommensurate filling\cite{we}, and found the transition always occurred when the fermion density was equal to or exceeded the boson density. 

Our aim here is to report the similar calculation in the other case, namely the case that the fermion density is below the boson density,
extending the phase diagram obtained in Ref. \citen{we} toward the negative side of $\delta\rho =(N_f-N_b)/N$, where $N_f$, $N_b$, and $N$ are the numbers of the fermions, the bosons and the lattice sites respectively.

We consider the mixture of $N_f$ spinless fermions and $N_b$ bosons on an $N$-site lattice by employing a Bose-Fermi Hubbard Hamiltonian:
\begin{eqnarray}
\mathcal{H}&=&-\sum_i\sum_{\alpha=f,b}\left[ t_{\alpha}(a_{\alpha,i}^\dagger 
a_{\alpha,i+1}+h.c.)+\mu_{\alpha} n_{\alpha,i} \right] \nonumber\\
& &{}+\sum_i \left[ \frac{U_{bb}}{2} n_{b,i}(n_{b,i}-1)+U_{fb}n_{f,i}n_{b,i} \right], \label{m1}
\end{eqnarray}
where $a_{\alpha,i}^\dagger$ and $a_{\alpha,i}$ are respectively creation and annihilation 
operators for fermions ($\alpha =f$) or bosons ($\alpha =b$) on the $i$-th site, and $n_{\alpha,i}=a_{\alpha,i}^\dagger a_{\alpha,i}$.
Hopping energy and chemical potential are denoted respectively by $t_{\alpha}$ and $\mu_{\alpha}$.
$U_{bb}$ and $U_{fb}$ are on-site boson-boson and fermion-boson repulsive interactions respectively. 
We set $t_f=t_b=1$ as an energy unit.

Reference \citen{Imambekov} proved that the system always stays in the mixing phase as far as the fermion-boson interaction has the same magnitude as the boson-boson interaction, and also we reported in our previous work\cite{we} that the transition to the demixing phase occurred when the on-site boson-boson interaction is sufficiently smaller than the on-site fermion-boson interaction.
In the present work, we thus set $U_{bb}=0$ to focus on how the transition point shifts with the change of $\delta\rho$.

To observe the mixing-demixing transition, we measured a correlation function defined by\cite{we},
\begin{equation}
C =\langle (n_{b,i-1}+n_{b,i+1})n_{b,i}\rangle - \langle 
(n_{f,i-1}+n_{f,i}+n_{f,i+1})n_{b,i}\rangle . \label{m2}
\end{equation}
When the two components are spatially mixed, the correlation function $C$ tends to be smaller, or negatively larger,
while it has a large positive value when they are demixed. Although we calculated other physical quantities, $C$ exhibited the change at the transition point most clearly and we decided to use $C$ to locate the transition point.

We performed Monte Carlo simulations with the world line algorithm on a 1D periodic lattice. 
Temperature was fixed to $T=0.08$, the Trotter decomposition number $L$ to $100$ and the number of sites $N$ to $30$.
In the simulations, we changed the number of the fermions and the bosons, $N_f$ and $N_b$ respectively, while fixing the total number $N_f+N_b=14$. As stated in the above
we are interested in the case $N_f < N_b$. 
\begin{figure}[tb]
\begin{tabular}{cc}
\resizebox{33mm}{!}{\includegraphics{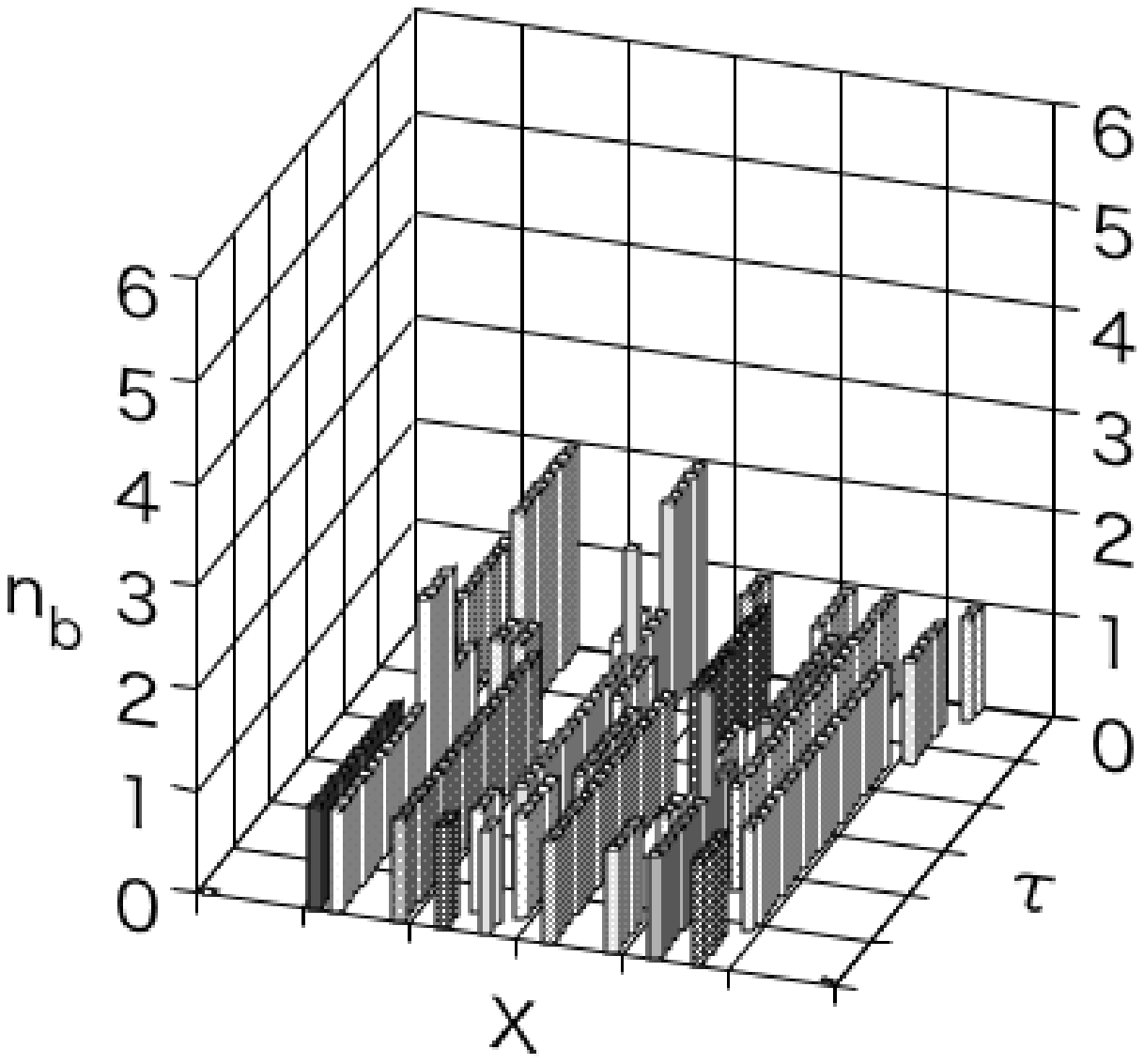}} & 
\resizebox{33mm}{!}{\includegraphics{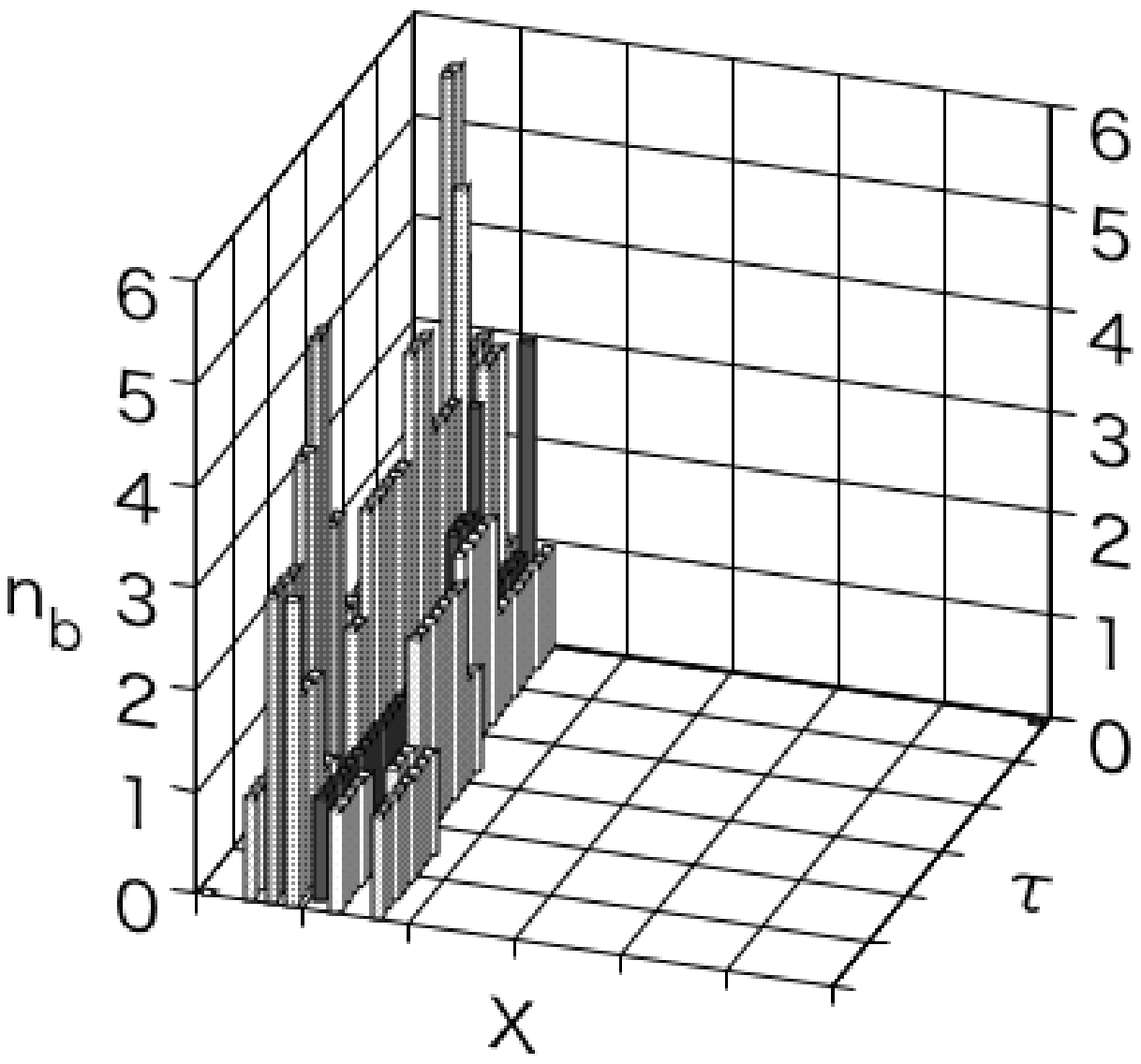}}\\
\resizebox{33mm}{!}{\includegraphics{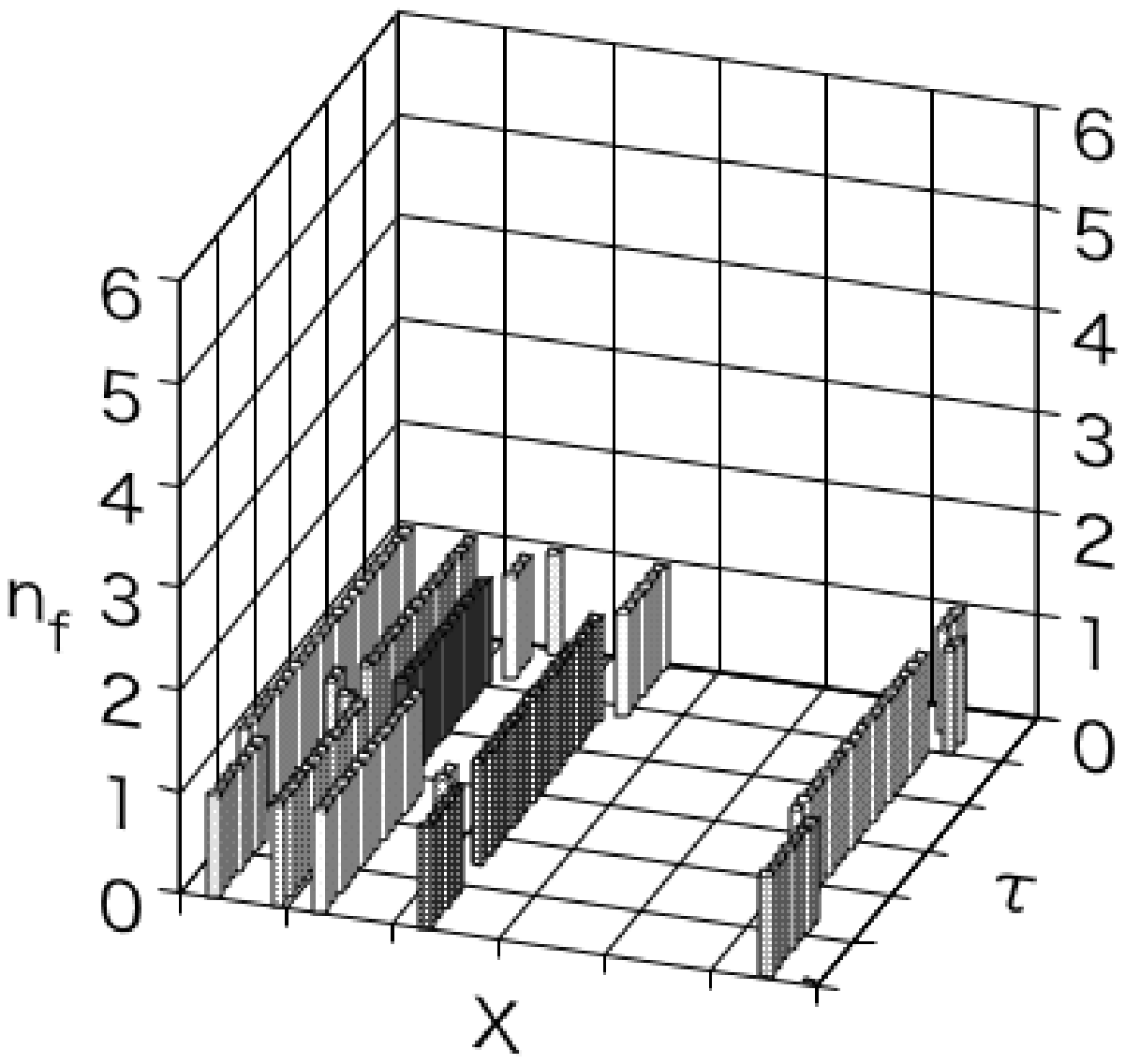}} & 
\resizebox{33mm}{!}{\includegraphics{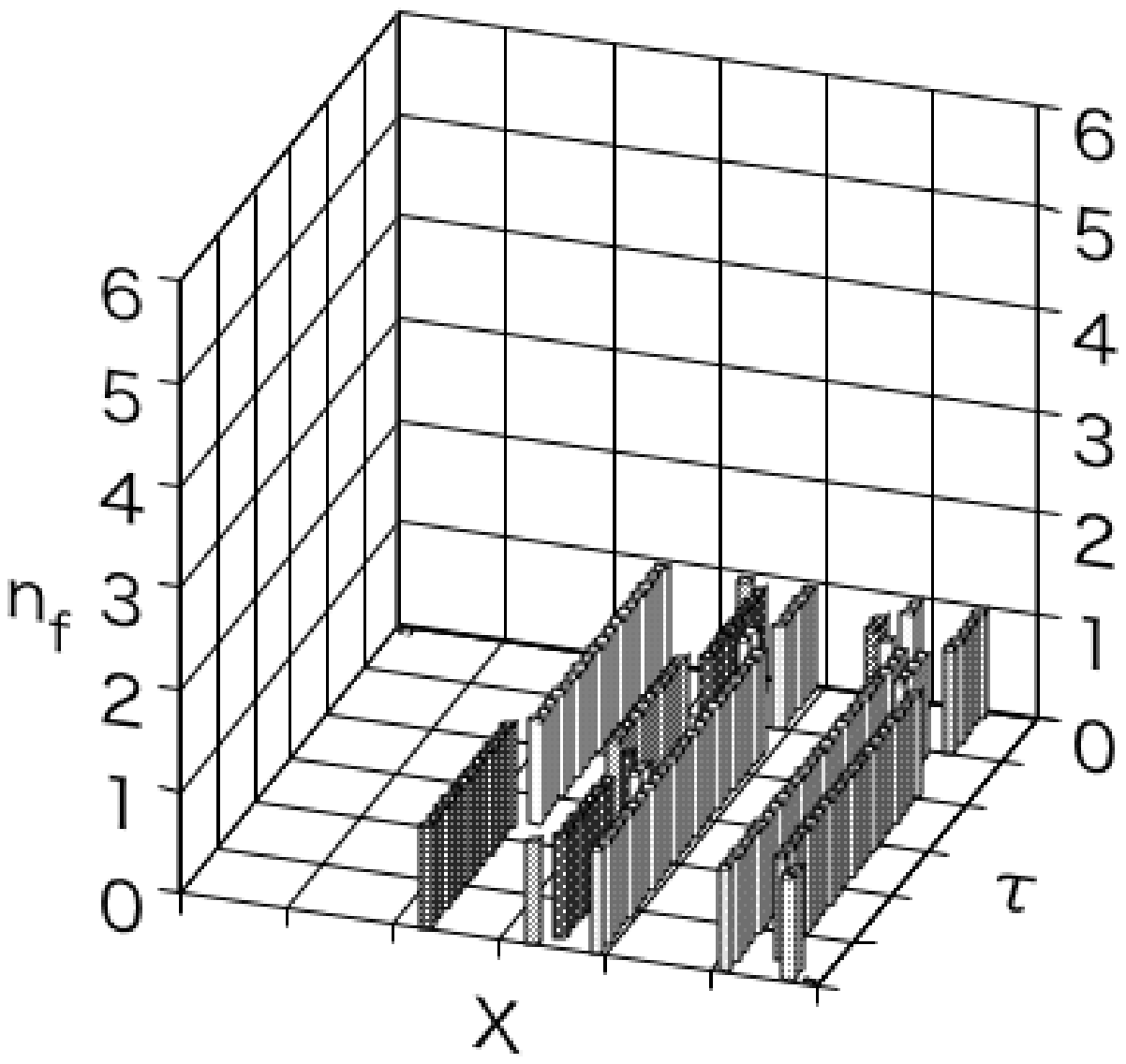}}\\
(a) & (b)
\end{tabular}
\caption{The snapshot of the boson configuration (upper) and the fermion configuration (lower) with $\delta\rho = -4/30$ and $U_{fb} = 0.4$ (a) and $4$ (b).  $X$ is the real space axis and $\tau$ is the imaginary time axis. $n_{b(f)}$ shows the number of the bosons (fermions) on each site.}
\label{f1}
\end{figure}

At first, we show the snapshot of typical configurations for two components in Fig. \ref{f1}, where the upper figures show the boson configurations and the lower show the fermion configurations when $\delta\rho = -4/30$ and $U_{fb}=0.4$ in (a) and $4$ in (b). With rather weak fermion-boson interactions, the system stays in the mixing phase as shown in (a), while the strong interactions drive the system into the demixing phase as in (b). 
\begin{figure}[tb]
\begin{tabular}{cc}
\resizebox{40mm}{!}{\includegraphics{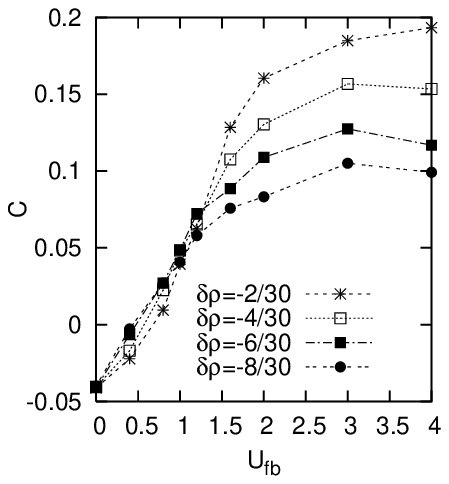}} & 
\resizebox{40mm}{!}{\includegraphics{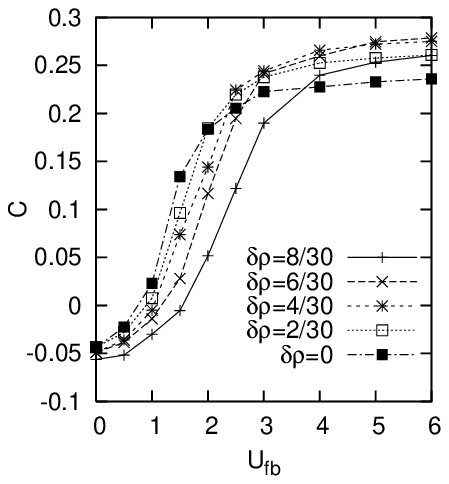}}\\
(a) & (b)
\end{tabular}
\caption{The correlation function $C$ as a function of $U_{fb}$ in the cases of $N_f < N_b$ ($\delta \rho < 0$) (a) and $N_f > N_b$ ($\delta \rho > 0$) (b). The result in (b) was obtained in our previous paper\cite{we} and used here for comparison.}
\label{f2}
\end{figure}

The transition can be seen more clearly in viewing the correlation function $C$, which is shown in Fig. \ref{f2} as a function of the
fermion-boson interaction $U_{fb}$. Figure \ref{f2}(a) presents the correlation function behavior at $\delta\rho =-2/30, -4/30, -6/30,$ and  $-8/30$. 
For comparison, we also show the previous result at $\delta \rho =0, 2/30, 4/30, 6/30,$ and $8/30$. 
As seen in the figure, there is a characteristic uprise in every curve of (a) and (b). 
This feature can be regarded as the occurrence of the transition from the mixing phase to the demixing at not only $\delta \rho \ge 0$ but also $\delta \rho <0$. It is interesting to note that the transition point seems to shift to the smaller $U_{fb}$ side as the number density of the fermions decreased relatively against that of the bosons. (Remember that the total number of the fermions and bosons was fixed in the simulation.) In other words, the shift of the transition point is not symmetric between the positive and the negative sides of $\delta\rho$, but it is rather monotonic as $\delta\rho$ changes from the positive to the negative value. 

\begin{figure}[tb]
\begin{center}
\resizebox{48mm}{!}{\includegraphics{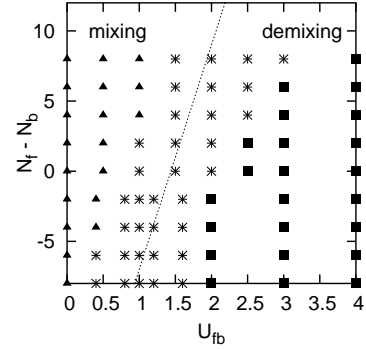}}
\end{center}
\caption{Phase diagram in $N_f -N_b$ vs $U_{fb}$. The triangle symbols and the square symbols present the mixing states and the demixng states, respectively. The asterisks are marked for undetermined phase points. The points of $\delta \rho >0$ region were obtained in our previous paper\cite{we}. The dotted line is just the guides for eyes.}
\label{f3}
\end{figure}

Our final aim of this short note is to complete the phase diagram in terms of the mixing and demixing phases. The phases could be determined by checking the snapshots and the behavior of the correlation function $C$, although the somewhat broad transition curves of $C$ did not allow us to pinpoint the transition point.
We therefore classified each parameter point into three states: Mixing, demixing, and undetermined. The last one corresponds to the uprise part of the correlation function, i.e. the transition area (not "point"). The broad transition curves are caused by the finite size effect of the system, which we checked by changing the system size.

Fig. \ref{f3} shows the phase diagram in the $\delta\rho - U_{fb}$ plane, combined with the one in the area $N_f >N_b$ obtained in the previous work\cite{we}. 
The triangle symbols and square symbols denote the mixing states and the demixing states, respectively. The asterisks represent the transition area.
Although we did not identify clearly the position of the transition point, the figure strongly suggests that the transition point would shift to the smaller $U_{fb}$ side as $\delta \rho$ changes from the positive to the negative value. 
This behavior can be understood in the following way. Making a cluster in the demixing state is unfavorable in terms of the kinetic energy of the fermions, because the fermions cannot move around in the dense cluster. Therefore the demixing state requires sufficiently large $U_{fb}$ to compensate the loss in the kinetic energy. However, as $\delta\rho$ changes from the positive to the negative value, i.e. as the number of the fermions decreases, the loss in the fermion kinetic energy becomes smaller, which makes it easier for the system to undergo the transition to the demixing state. This explains the shift of the transition point.

In conclusion, we numerically studied the mixing-demixing transition of 1D boson-fermion mixtures at low fermion densities with the total particle density fixed, and obtained
the phase diagram in the $\delta\rho - U_{fb}$ plane. The phase diagram suggests that the
qualitative feature of the system is not determined simply by the magnitude of the difference in the number density between the fermions and the bosons.

The authors would like to thank Prof. Jo and Prof. Oguchi for their continuous support.

\end{document}